\documentclass[preprint,12pt]{elsarticle}

\pdfoutput=1
\usepackage{ulem}
\usepackage{amsfonts}
\usepackage{mathrsfs}
\usepackage{amsthm}
\usepackage{multirow}
\usepackage{bbding}
\usepackage{amssymb}
\usepackage{amsmath}
\usepackage{graphicx,color}
\usepackage{mathrsfs}
\usepackage{rotating}
\usepackage{algorithm}
\usepackage{algorithmic}

\journal{Physica A}

\begin{document}

\begin{frontmatter}



\title{On Equivalence of Likelihood Maximization of Stochastic Block Model and Constrained Nonnegative Matrix Factorization}

\tnotetext[mytitlenote]{This work was supported by Program for Innovation Research in Central University of Finance and Economics and Young Elite Teacher Project of Central University of Finance and Economics.}
\author{Zhong-Yuan Zhang\corref{mycorrespondingauthor}}
\cortext[mycorrespondingauthor]{Corresponding author.}
\ead{zhyuanzh@gmail.com}

\author{Yujie Gai}
\author{Yu-Fei Wang}
\author{Hui-Min Cheng}
\author{Xin Liu}
\address{School of Statistics and Mathematics, Central University of Finance and Economics, Haidian District, Beijing 100081, China}

\begin{abstract}
Community structures detection in complex network is important for understanding not only the topological structures of the network, but also the functions of it. Stochastic block model and nonnegative matrix factorization are two widely used methods for community detection, which are proposed from different perspectives. In this paper, the relations between them are studied. The logarithm of likelihood function for stochastic block model can be reformulated under the framework of nonnegative matrix factorization. Besides the model equivalence, the algorithms employed by the two methods are different.
Preliminary numerical experiments are carried out to compare the behaviors of the algorithms. 
\end{abstract}

% \begin{keyword}



% \end{keyword}

\end{frontmatter}


\section{\label{Introduction}Introduction}
One of the fundamental problems in network analysis is community structures
detection (\cite{girvan2002community,newman2004finding}). In most cases, a community is  a group of nodes which connect with each other tightly while connect loosely with
 the rest of the network. There are other types of communities as well.  Communities often correspond to functional
units. For example, in a social network, a community might
correspond to  a group of people brought together by a common interest. Detection of communities is very important
for understanding not only the topological structures of the network, but also the functions of it, such as how the nodes communicate with each other,
or how new ideas diffuse in the network (\cite{nematzadeh2014optimal}), etc.

Many  kinds of methods have been proposed to detect community structures in the literature recently. One group of methods is  based on generative model.
The basic motivation is that the network we observed is an instance generated by a set of hidden parameters,
 and we can detect the community structures in the network by revealing the parameters.
Among them,  the most
representative one is the stochastic block model (SBM, \cite{holland1983stochastic, snijders1997estimation,nowicki2001estimation,newman2007mixture,goldenberg2010survey,airoldi2009mixed,bickel2009nonparametric,
newman2012communities,karrer2011stochastic,decelle2011asymptotic}).
SBM provides a well-founded principled approach for understanding the network structures, and is very flexible such that not only the traditional type of community structures, but also a wide variety of structures in networks can be formulated into the model and be detected (\cite{newman2012communities,aicher2014learning}). In this paper, our focus is on the stochastic block models proposed by Karrer and Newman (\cite{karrer2011stochastic}), and its variants (\cite{shen2011exploring, chen2014overlapping,larremore2014efficiently,aicher2014learning}).
 %Recently,  many other kinds of block models  were introduced in the  literatures, such as degree-corrected SBM (\cite{karrer2011stochastic}), mixture models for
% directed networks (\cite{shen2011exploring}),  mixed membership SBM for modeling
% overlapping communities (\cite{airoldi2009mixed}) and signed networks (\cite{chen2014overlapping}).
%From the algorithmic point of view, such  methods lead to criteria to be optimized over all partitions,
%such as the profile likelihood under the assumed model.

%(Robust Kernel Nonnegative Matrix Factorization, Zhichen Xia Chris Ding Edmond)
Another group of methods for community detection is based on  optimization of some
 global criteria over all possible network partitions,  including
graph partitioning (\cite{schaeffer2007graph}), spectral clustering(\cite{krzakala2013spectral}),   modularity maximization (\cite{newman2004finding,newman2006modularity}) and  nonnegative  matrix mactorization(\cite{lee1999learning}), etc.
Nonnegative matrix factorization (NMF) was originally
proposed as a method for finding matrix factors with parts-of-whole interpretations (\cite{lee1999learning, lee2001algorithms}), and has become a powerful tool for data analysis with enhanced interpretability. By accommodating a variety of objective functions,  NMF has been successfully applied to a lot of distinct areas (\cite{li2001learning,cooper2002summarizing, brunet2004metagenes,pauca2004text, sra2005generalized,shashua2005non}).
  Specifically, NMF has been successfully adopted to community structures detection recently (\cite{zarei2009detecting,psorakis2011overlapping,zhang2013overlapping,
  cao2013identifying,zhang2015community}). Both spectral clustering and probabilistic latent semantic indexing can be reformulated under the framework of NMF (\cite{ding2005equivalence,ding2008equivalence}).

Although there were  so many crucial researches of community detection  based on SBM and NMF respectively, few  attempt has been made to build the connections
between them while emphasizing their differences in community detection.
In this paper we prove that the likelihood functions to be maximized for SBM
are equivalent to the objective functions for nonnegative matrix factorization model with constraints.
% the standard stochastic block models and its several variants can be reformulated under the NMF framework.
To the best of our knowledge, this is the first time establishing the connections between SBM and NMF.
% Specifically, we  show that
Empirical experiments are carried out to compare the difference between the algorithms employed by the two models.

The rest of the paper is organized as follows. From Sect. \ref{standard} to Sect. \ref{signed}, we give the connections between six types of block models and nonnegative matrix factorizations, respectively. The six block models are the standard SBM, the degree-corrected SBM, the bipartite SBM, the normal distributed SBM, the directed SBM, and the signed SBM. Experimental results are given in Sect. \ref{experi}. And Sect. \ref{conclu} concludes.

\section{Equivalence  of likelihood maximization of standard stochastic block model and constrained nonnegative matrix factorization}\label{standard}
Let $\mathscr{G}$ be an undirected multigraph with $n$ vertices, possibly including self-edges.  Each vertex $i$ belongs to one of $c$ latent communities and assume the number of edges between each two nodes (or self-edges)  to be independently Poisson distributed.
 Let  $A$ be the adjacency matrix of $\mathscr{G}$, with  its  element   $A_{ij}$ denoting the number of edges between vertices $i$ and $j$, and $W$ be a $c\times c $ matrix with its   element   $w_{rs}$  the expected value of $A_{ij}$  for vertices $i$ and $j$ lying in community $r$ and $s$ respectively.
Also, we introduce  community membership  matrix $G$ to record the community assignment of vertices in the network $\mathscr{G}$,  where $G_{ir}=1$ if vertex $i$ belongs to community $r$; 0  otherwise.

According to Karrer and  Newman (\cite{karrer2011stochastic}), the probability $P(\mathscr{G}| W,G)$ of graph $\mathscr{G}$  given the parameters $W$ and the community assignment matrix $G$ is
 \begin{align}\label{peq1}
   P(\mathscr{G}| W,G)= \frac{1}{\prod_{i<j}}A_{ij}!\prod_i 2^{A_{ii}/2}(A_{ii}/2)!\times \prod_{r,\,s}w_{rs}^{m_{rs}/2}\exp(-\frac{1}{2}n_r n_s w_{rs}),
\end{align}
where  $n_r$ is the number of vertices in community $r$ and $\displaystyle\sum_{r=1}^c n_r=n$, $m_{rs}$ is the total number of edges between community $r$ and $s$, or twice  that number if $r=s$ and $\displaystyle\sum_{k=1}^c G_{ik}=1$.
The goal is to maximize this probability (\ref{peq1}) with respect to the unknown model parameters $W$ and the community membership matrix $G$. By neglecting constants which are independent of the parameter $W$ and the community assignment matrix $G$, the problem above is reduced to
\begin{align*}%\label{4eq1}
  \mathop{\max}_{W,\,G }\log P(\mathscr{G}| W,G)= \mathop{\max}_{W,\,G } \sum_{r,\,s} (m_{rs}\log w_{rs}-n_r n_s w_{rs}),
\end{align*}
which  is equivalent to
\begin{align}\label{4eq2}
\mathop{\min}_{W,\,G } \sum_{r,\,s}- \left(m_{rs}\log{ w_{rs}}- n_r n_s w_{rs}\right ).
\end{align}
%Now we  will provide the equivalent nonnegative matrix factorization form of (\ref{4eq2}).
% To do so, we will do some transformations for  equation ({\ref{4eq2}}) first.

The first  term in equation ({\ref{4eq2}}) is
\begin{eqnarray*}
   -\sum_{r,\,s} m_{rs}\log { w_{rs}}
   &=&-\sum_{r,\,s}\sum_{i,\,j} A_{ij}G_{ir}G_{js}\log{w_{rs}}\\
    &=&-\sum_{i,\,j} A_{ij}(\sum_{r}G_{ir}\sum_{s}G_{js}\log{w_{rs}})\\
  %&=&-\sum_{i,j} A_{ij}(\sum_r G_{ir}G_{js^*}\log{w_{rs^*}})\label{4eq2-1}\\
  &=&-\sum_{i,\,j} A_{ij}( G_{ir^*}G_{js^*}\log{w_{r^*s^*}})\\
  &=&-\sum_{i,\,j} A_{ij}(\log( G_{ir^*}G_{js^*}{w_{r^*s^*}}))\\
  &=&-\sum_{i,\,j} A_{ij}\log(\sum_{r,s}G_{ir}G_{js}{w_{rs}})\\
   &=&-\sum_{i,\,j}A_{ij}\log {(G W G^T)_{ij}}\\
    &=&\sum_{i,\,j}A_{ij}\log {\frac{1}{(G W G^T)_{ij}}},
  \end{eqnarray*}
the third equality above holds because there is only one 1 in each row of matrix $G$, which is denoted by $G_{ir^*}$ and $G_{js^*}$, respectively.

The second term in equation (\ref{4eq2}) is
\begin{align}\label{2eq1}
  \sum_{r,\,s} n_r n_s w_{rs}= \textbf{n}^T W \textbf{n}
  = \textbf{1}^T G W G^T \textbf{1}=\sum_{i,\,j}(G W G^T)_{ij},
 \end{align}
where   $\textbf{n}=(n_1, \cdots, n_c)^T$ is  a $c\times 1$ column vector,  and
 $\textbf{1}$ is a $c\times 1$ vector with all elements 1.  $A^T$ is the transpose   of a matrix  (or a vector) $A$.
%Then we have
%\begin{eqnarray}\label{2eq2}
%-\sum_{r,s} m_{rs}\log { w_{rs}}=\sum_{i,j}A_{ij}\log\frac{1} {(G W G^T)_{ij}}
%\end{eqnarray}
%The same reason for  the  equality  (2.9).

 Hence,  the optimal problem (\ref{4eq2}) is equivalent to
\begin{align*}
  \mathop{\min}_{W,\,G }\left[\sum_{i,\,j}A_{ij}\log \frac{1}{(G W G^T)_{ij}}+\sum_{i,\,j}(G W G^T)_{ij}\right].
\end{align*}

By adding some constants which are independent of $W$ and $G$, we have the following binary matrix factorization problem
\begin{equation}\label{2eq2}
\left\{
\begin{array}{rl}
 \displaystyle\min_{W,\,G } & \sum\limits_{i,\,j}\left[A_{ij}\log \displaystyle\frac{A_{ij}}{(G W G^T)_{ij}}+(G WG^T)_{i,j}-A_{ij}\right]\\\\
 s.t. &\displaystyle\sum_{r=1}^c G_{ir}=1, ~ i =1,2, \cdots, n,\\\\
  & G_{ir} = 0\, \mbox{or}\, 1, i =1,2, \cdots, n;~ r=1,2,\cdots, c, \\\\
  & w_{rs}\geq 0, i =1,2, \cdots, n;~ r,s=1,2,\cdots, c.
\end{array}
\right.
\end{equation}
which is typically a constrained nonnegative matrix factorization model, and can be naturally extended to overlapping community structures detection by relaxing the constraints on $G$ from binary to non-negativity: $\displaystyle\sum_{r=1}^c G_{ir}=1, ~ i =1,2, \cdots, n,$ $ G_{ir}\geq 0, i =1,2, \cdots, n;~ r=1,2,\cdots, c.$

In the following sections from Sect. \ref{correlated} to Sect. \ref{normal}, without loss of generality, we wrote out the relaxed nonnegative matrix factorization model directly. Note that although the relaxed NMF model is an approximation, it outputs the same results for nonoverlapping community detection problem by discretizing $G$ \cite{zhang2013overlapping}.

\section{Equivalence  of likelihood maximization of degree-corrected stochastic block model and constrained nonnegative matrix factorization}\label{correlated}
 In the standard SBM mentioned in Sect. \ref{standard},
  vertices in the same
community are identical, i.e.,  vertices in the same community have equal
probability connecting to others,
and thus are supposed to have the same degree distribution, which is not realistic, since real networks are often degree heterogeneity.
 Karrer and  Newman (\cite{karrer2011stochastic}) proposed the degree-corrected SBM to  take  degree heterogeneity into account when generating a network.
   In this section, we will prove the equivalence of degree-corrected SBM and NMF.

When considering  the degree-corrected SBM, the generation  of   $\mathscr{G}$ depends not only on the parameters introduced previously, but also on a new set of parameters $\theta_i~, i =1,2, \cdots, n$, which is the degree weight of vertex $i$ satisfying the condition $\displaystyle\sum_{i=1}^n\theta_iG_{ir}=1,~ r=1,2,\cdots,c$. The expected value of the  edge number between two vertices $i$ and $j$ is $\theta_i\theta_j w_{rs}$, instead of $ w_{rs}$,  where $r, s$ are the communities that vertex $i$ and $j$ belong to respectively.

   For the degree-corrected SBM, the goal is to maximize the probability (\ref{de-p1}) below with respect to the unknown parameters $\theta$, $W$ and the community memebership matrix $G$, where  ${\theta}=(\theta_1,\cdots,\theta_n)^T$:
\begin{eqnarray}\label{de-p1}
  P(\mathscr{G}|{\theta},W,G)&=&\frac{1}{\prod_{i<j}A_{ij}!\prod_i 2^{A_{ii}/2}(A_{ii}/2)!}\nonumber\\
   && \times\prod_{i<j}(\theta_i\theta_j)^{A_{ij}}\prod_i (\theta_i^2)^{A_{ii}/2} \prod_{r,\,s}w_{rs}^{m_{rs}/2}\exp(-\frac{1}{2}w_{rs}),
\end{eqnarray}
or
\begin{eqnarray}\label{de-p2}
  &&\log P(\mathscr{G}|{\theta},W,G)\nonumber\\&=& 2\sum_{i<j}A_{ij}\log(\theta_i\theta_j)+2\sum_i \frac{A_{ii}}{2}\log \theta_i^2+
  \sum_{r,\,s}(m_{rs}\log w_{rs}-w_{rs})\nonumber\\
  &=& \sum_{i,\,j}A_{ij}\log(\theta_i\theta_j)+\sum_{r,\,s}(m_{rs}\log w_{rs}-w_{rs}).
\end{eqnarray}

 Hence the problem is
\begin{align}\label{4eq7}
  \mathop{\max}_{{\theta},\,W,\,G}\log P(\mathscr{G}|{\theta}, W,G)=\mathop{\max}_{{\theta},\,W,\,G} \left[\sum_{i,\,j}A_{ij}\log{(\theta_i\theta_j)}+\sum_{r,\,s}\left(m_{rs}\log w_{rs}-w_{rs}\right)\right],
\end{align}
which is equivalent to
\begin{align}\label{4eG8}
  \mathop{\min}_{{\theta},\,W,\,G} \left(\sum_{i,\,j}A_{ij}\log\frac{1}{\theta_i\theta_j}+\sum_{r,\,s}m_{rs}\log \frac{1}{w_{rs}}+\sum_{r,\,s}w_{rs}\right).
\end{align}

The first two terms in (\ref{4eG8}) can be combined together as
\begin{eqnarray}\label{4eG9}
 &&\sum_{i,\,j} A_{ij}\log\frac{1}{\theta_i\theta_j}+\sum_{r,\,s}m_{rs}\log \frac{1}{w_{rs}}\nonumber\\
 &=&\sum_{i,\,j}A_{ij}\log\frac{1}{\theta_i\theta_j}+\sum_{i,\,j}A_{ij}\log \frac{1}{(G W G^T)_{ij}}\nonumber \\
 &=&\sum_{i,\,j}A_{ij}\log\frac{1}{\theta_i\theta_j(G W G^T)_{ij}}\nonumber\\
 &=& \sum_{i,\,j}A_{ij}\log\frac{1}{(({\theta}{\theta}^T) \otimes (G W G^T))_{ij}}.
\end{eqnarray}

The third term in (\ref{4eG8}) is
\begin{eqnarray}\label{4eG10}
  \sum_{r,\,s}w_{rs}&=&\mathbf{1}^T  W \mathbf{1}\nonumber\\
  &=&{\theta}^TG W G^T {\theta}\nonumber\\
  &=& \sum_{i,\,j} \theta_i \theta_j (G W G^T)_{ij}\nonumber\\
    &=& \sum_{i,\,j}  \left(({\theta}{\theta}^T) \otimes (G W G^T)\right)_{ij},
\end{eqnarray}
 where $\otimes$  denotes the  dot product of two matrices (or vectors) with the same dimensions. And  the second equality above  holds because  sum of weights in one community equals 1, that is,
 $\displaystyle\sum_{i=1}^n\theta_iG_{ir}=1,~ r=1,2,\cdots,c.$

By combining (\ref{4eG9}) and (\ref{4eG10}) together,  the optimal problem (\ref{4eq7}) is then equivalent to
$$\hspace{-3mm}\left\{
\begin{array}{rl}
 \displaystyle\min_{\theta,\,W,\,G } & \displaystyle\sum_{i,\,j}\left[ A_{ij}\log \frac{A_{ij}}{(({\theta}{\theta}^T) \otimes (G W G^T))_{ij}}+(({\theta}{\theta}^T) \otimes (G W G^T))_{ij}-A_{ij}\right]\\\\
 s.t. &\displaystyle\sum_{i=1}^n \theta_i G_{ir}=1,~~~~~ r=1,2, \cdots, c.\\\\
  & \displaystyle\sum_{r=1}^c G_{ir}=1, ~~~~~~~ i =1,2, \cdots, n.\\\\
  & G_{ir},\theta_i,w_{rs}\geq 0, ~~~i =1,2, \cdots, n;~ r,s=1,2,\cdots, c.
\end{array}
\right.$$

That is,  optimization problem over the degree-corrected SBM  is  equivalent to the weighted NMF model.

%According to (\textbf{ Shen, Cheng, and Guo, 2011. Exploring the structural regularities in  networks.})
\section{Equivalence of likelihood maximization of bipartite stochastic block model and constrained nonnegative matrix factorization}\label{bipartite}
In this section, we consider the equivalence of  bipartite SBM and NMF. $\mathscr{G}$ is an undirected bipartite multigraph with $n$ vertices, possibly including self-edges. There are two types of vertices, i.e. type I and type II, and only vertices of different types may be connected. Each community contains vertices of a single type.  The number of vertices in type I and type II are $n_1$ and $n_2$, respectively. Let $B$ be a $n_1\times n_2$ bipartite adjacency matrix with $B_{ij}=1$ if there is an edge between $i$ and $j$ from type I and type II respectively; $0$ otherwise, and  $A$ be a $n\times n$ adjacency matrix related to $B$ as
\begin{align*}
A=\left(
                      \begin{array}{cc}
                      0 &  B \\
                      B^T & 0 \\
                      \end{array}
                    \right).
\end{align*}
We assume that the number of edges between each pair of vertices (including self-edges) is independently Poisson distributed, similar to that in Sect. \ref{standard},  and define $w_{rs}$ to be the expected value of $A_{ij}$  for vertices $i$ and $j$ lying in community $r$ and $s$ respectively. Since vertices of the same type cannot be connected, we have
\begin{align*}
  w_{rs}=0 ~~ \textrm{when } ~~ T_{rs}=0,
\end{align*}
where $T$ is a $c \times c$ matrix with its element $T_{rs}=1$ if the types of community $r$ and $s$ are different; 0 otherwise. Other notations are defined in Sect. 2.

  According to Larremore {\it  et~ al.}  (\cite{larremore2014efficiently}), the probability of the network  $\mathscr{G}$ is that,
 \begin{align}\label{bieq1}
  P(\mathscr{G}| W,G, T)= \prod_{i<j, i\in r, j\in s,T_{rs}=1}
\frac{( w_{rs})^{A_{ij}} } { A_{ij}! }
\exp(- w_{rs}),
\end{align}
where $i\in r$ stands for vertex $i$  belonging to community $r$.
After a small amount of manipulation, and neglecting constants, taking the logarithm, (\ref{bieq1}) is equivalent to
 \begin{align}\label{bieq2}
  \log P(\mathscr{G}| W,G, T)=  \sum_{r,\,s,~T_{rs}=1}\left(m_{rs}\log w_{rs}-n_r n_s w_{rs}\right).
\end{align}
The goal is to maximize (\ref{bieq2}) with respect to  $W$,  $G$ and $T$.

Since
\begin{eqnarray*}
  \sum_{r,\,s,~ T_{rs}=1}n_r n_s w_{rs}=\textbf{1}^T G (T\otimes W) G^T \textbf{1},
\end{eqnarray*}
and
\begin{eqnarray*}
   \sum_{r,\,s,~T_{rs}=1} m_{rs}\log { w_{rs}}&=&-\sum_{r,\,s~T_{rs}=1} m_{rs}\log \frac{1}{ w_{rs}}\\&=&-\sum_{r,\,s~T_{rs}=1}\sum_{i,\,j} A_{ij}G_{ir}G_{js}\log\frac{1}{w_{rs}}\\
&=&-\sum_{r,\,s}T_{rs}\left(\sum_{i,\,j} A_{ij}G_{ir}G_{js}\right)\log\frac{1}{({T\otimes W})_{rs}}\\
&=&-\sum_{i,\,j} A_{ij}\left(\sum_{r,\,s}T_{rs}G_{ir}G_{js}\log\frac{1}{({T\otimes W})_{rs}}\right)\\
  &=&-\sum_{i,\,j}A_{ij} \log\frac{T_{r^*s^*}G_{ir^*}G_{js^*}}{(T\otimes W)_{r^*s^*}}\\
&=& \sum_{i,\,j}A_{ij}\log{\sum_{r,\,s}\left( T_{rs}G_{ir}G_{js}(T\otimes W)_{rs}\right)}\\
&=& \sum_{i,\,j,\,r,\,s}A_{ij}\log\sum_{r,\,s}\left(G_{ir}G_{js}(T\otimes W)_{rs}\right)\\
&=&\sum_{i,\,j} A_{ij} \log{\left(G(T\otimes W)G^T \right)_{ij}},
  \end{eqnarray*}
the maximization of expression (\ref{bieq2}) is equivalent to the following weighted nonnegative matrix factorization model:
$$\left\{
\begin{array}{rl}
 \displaystyle\min_{H,\, W} & \displaystyle\sum_{i,\,j}\left[ A_{ij} \log\frac{A_{ij}}{\left(G(T\otimes W)G^T\right)}_{ij}+{\left(G(T\otimes W)G^T\right)}_{ij}-A_{ij}\right]\\\\
 s.t. &\displaystyle\sum_{r=1}^c G_{ir}=1, ~ i =1,2, \cdots, n.\\\\
  & G_{ir}, w_{rs}\geq 0, ~i =1,2, \cdots, n;~ r, s=1,2, \cdots, c.\\\\
  & T_{rs} = 0, 1, r,s=1,2,\cdots,c.
\end{array}
\right.$$

Similarly, for the bipartite degree-corrected SBM,  it's easy to obtain the equivalence to NMF, too.
\section{Equivalence of likelihood maximization of normal distributed edge-weighted stochastic block model and constrained nonnegative matrix factorization }\label{normal}
Aicher, Jacobs and Clauset (\cite{aicher2014learning}) studied the weighted stochastic block model, and the normal distributed  edge weight, as one case, was provided. In this section, we will prove that it is also equivalent to NMF. The model is defined as follows.  For an undirected multigraph $\mathscr{G}$ on $n$ vertices, the weight of edge between two vertices $i, j$ is supposed to drawn from a normal distribution  $N(\mu_{rs}, \sigma^2_{rs})$,  where $r, s$ are the community assignments of vertices $i$ and $j$ respectively. In this case, the likelihood function is
\begin{eqnarray}\label{bieq3}
    P(\mathscr{G}| \mu,\sigma^2)=\prod_{i,\,j~ i\in r,\, j\in s} \frac{1}{\sqrt{2\pi}\sigma_{rs}} \exp\left(-\frac{(A_{ij}-\mu_{rs})^2}{2\sigma^2_{rs}}\right),
\end{eqnarray}
with $i\in r$ standing for vertex $i$ lying in community $r$.
The goal is to maximize (\ref{bieq3}) with respect to parameters $\mu$ and $\sigma$.  Fixing $\sigma$, the maximization of (\ref{bieq3}) is equivalent to minimizing the following expression over $\mu_{rs}$,
\begin{eqnarray}\label{bieq4}
\sum_{i,\,j~ i\in r,\, j\in s} (A_{ij}-\mu_{rs})^2.
\end{eqnarray}
Meanwhile, note that the expectation
\begin{eqnarray}\label{bieq5}
\mu_{rs}=\sum_{r,\,s}G_{ir}G_{js}w_{rs}=(GWG^T)_{ij},
\end{eqnarray}
where $w_{rs}$ is the expected edge weight between two vertices belonging to community $r$ and $s$, respectively, and $G$ again is the community membership matrix with
 $G_{ir}=1$ if vertex $i$ belongs to group $r$; 0 otherwise.  The equation (\ref{bieq5}) is very critical, because it reveals the relationship between parameter $ \mu_{rs}$ and the community assignment $G$.

  Then the optimization problem (\ref{bieq4}) is reduced to the following NMF model:
\begin{equation}\label{normal1}
\left\{
\begin{array}{rl}
 \displaystyle\min_{G,\, W} & \displaystyle\sum_{i,\,j}(A_{ij}-(GWG^T)_{ij})^2\\\\
 s.t. &\displaystyle\sum_{r=1}^c G_{ir}=1, ~ i =1,2, \cdots, n.\\\\
  & G_{ir}, w_{rs}\geq 0, ~i =1,2, \cdots, n;~ r, s=1,2, \cdots, c.
\end{array}
\right.
\end{equation}
 \section{Equivalence of likelihood maximization of directed stochastic block model and constrained nonnegative matrix factorization}\label{directed}
 %Another generalization of standard block model is directed block model.
  In this section, we turn to consider the directed SBM setting. In a directed SBM,  an edge is an ordered pair of  vertices,    that is, an edge from vertex $i$ to $j$  is different with  an edge from vertex $j$ to $i$.  This differs from the standard (undirected) SBM introduced in Sect. \ref{standard}, in that the latter is defined in terms of unordered pairs of vertices.

 Notations given in  Sect. \ref{standard} and Sect. \ref{correlated}  have to be redefined and some new notations are provided  here.  $\mathscr{G}$ is a directed graph with $n$ vertices.        $A$ is still an $n\times $n adjacency matrix  but is not symmetric,   $A_{ij}=1$  if there is an edge from node $i$ to node $j$ and 0 otherwise,  where $i$ is named tail node and $j$ is named head node.   For weighted networks, $A_{ij}$ is generalized to represent the weight of the edge from $i$ to $j$. $W$ is a $c\times c$ matrix with $w_{rs}$ denoting the probability that a randomly selected edge, of which the tail node  is from group $r$ and the head node is from group $s$.
 $F$ and $H$ are the community membership matrices, where ${F}$ is a $n\times c$ matrix with its element  $F_{ir}$ denoting the probability that the tail node $i$ is from community $r$, and $H$ is a $n\times c$ matrix with its element  $H_{js} $ denoting the probability that the head node $j$ is from $s$, respectively.

According to Shen, Cheng and Guo (\cite{shen2011exploring}),
 the goal is to maximize the probability $P(\mathscr{G}|{F},W,{H})$, which is the profile likelihood of the observed network,   or equally to maximize the logarithm of the probability,  with respect to the parameter $W$, and the community membership matrices $F$ and $H$. That is,
\begin{eqnarray*}
 \max_{{F},\,W,\,{H}} \log P(\mathscr{G}|{F},W,{H})=\max_{{F},\,W,\,{H}}\sum_{i,\,j} A_{ij} \log \left( \sum_{r,\,s} w_{rs} F_{ir}H_{js}   \right),\nonumber
\end{eqnarray*}
 with respect to the constraints $ \displaystyle\sum_{r,\,s=1}^c w_{rs}=1,   \displaystyle\sum_{i=1}^nF_{ir}=1,   \displaystyle\sum_{j=1}^nH_{js}=1, ~ r,s=1,2,\cdots, c$ (\cite{shen2011exploring}),
which is  equivalent to

\begin{equation}\label{directed-2}
\left\{
\begin{array}{lll}\vspace{5mm}
 &&\min\limits_{{F},\,W,\,{H}}\ \displaystyle\sum_{i,\,j} A_{ij} \log \frac{1} {\displaystyle\sum_{r,\,s} w_{rs} F_{ir}H_{js}} \\
&&\textrm{s.t.} ~~\displaystyle\sum_{r,\,s=1}^c w_{rs}=1,    \sum_{i=1}^n F_{ir}=1,   \sum_{j=1}^n H_{js}=1, ~ r,s=1,2,\cdots, c.
\end{array}
\right.
\end{equation}

Since
\begin{eqnarray*}
\sum_{r,\,s}w_{rs}F_{ir}H_{js}=\left(FWH^T\right)_{ij}
\end{eqnarray*}

 and
\begin{eqnarray*}
  \sum_{i,\,j}\sum_{r,\,s}w_{rs}F_{ir}H_{js}& = \displaystyle\sum_{i,\,r,\,s} w_{rs}F_{ir}\displaystyle\sum_j H_{js}\\
  &\hspace{-13mm}=\displaystyle\sum_{i,\,r,\,s} w_{rs}F_{ir}\nonumber\\
  &\hspace{-9mm}=\displaystyle\sum_{r,\,s}w_{rs}\displaystyle\sum_iF_{ir}\\
  &\hspace{-11mm}=\displaystyle\sum_{r,\,s}w_{rs}=1,
\end{eqnarray*}
 we have that the optimal problem (\ref{directed-2}) has the following equivalent non-negative matrix factorization form
$$\left\{
\begin{array}{rl}
 \displaystyle\min_{{F},\,W,\,{H}} &\left[ \displaystyle\sum_{i,\,j} A_{ij}\log\frac{A_{ij}}{\left( FWH^T\right)_{ij}}+\sum_{i,\,j}\left( FWH^T\right)_{ij}-\sum_{i,\,j}A_{ij}\right]\\\\
 s.t. &\displaystyle\sum_{r,\,s=1}^c w_{rs}=1,   \sum_{i=1}^n F_{ir}=1,   \sum_{j=1}^n H_{js}=1, ~ r,s=1,2,\cdots, c.\\\\
  & F_{ir}, H_{js}, w_{rs}\geq0, i, j=1,2,\cdots,n; r, s=1,2,\cdots,c.
\end{array}
\right.$$

which is actually also equivalent to probabilistic latent semantic indexing (\cite{ding2008equivalence}).
%where $\mathbf{1}_n$ is an $n\times 1$ column vector with all elements 1, similar to $\mathbf{1}_K.$

 \section{Equivalence of likelihood maximization of signed stochastic block model and constrained nonnegative matrix factorization}\label{signed}
 Networks possessing  both positive and negative links are called signed networks.
 In signed networks, most edges within a community  are positive links, and most edges across communities are negative links. Signed networks exist in many occasions.
 For example,
 in a social network,
positive links may denote  agreement  whereas negative links may denote
disagreement.   Chen {\it et al.} (\cite{chen2014overlapping}) studied the signed SBM. In this section, we will prove the equivalence of signed SBM and NMF.

Firstly, we update some notations. $\mathscr{G}$ is a signed network with $n$ vertices, and $A$ is the adjacency matrix.
We use $A^+$  and $A^-$  to denote the positive and
negative parts in the signed network, respectively.
That is,  $A_{ij}^+=A_{ij}$~ if~ $A_{ij}>0, $ 0 otherwise; $A_{ij}^-=-A_{ij}$~ if~ $A_{ij}<0,$   0 otherwise. Let ${H}$ be the community membership matrix with its element  $H_{ir}$ denoting the probability that the node $i$ is in the community $r$. $W$ is a $c\times c $ matrix with its element $w_{rs}$ denoting the probability of an edge choosing between community $r$ and $s$. The normalization constraints on $H$ and $W$ are $\displaystyle\sum_i H_{ir}=1$ and $\displaystyle\sum_{r,\,s}w_{rs}=1.$  Let $W_1=diag(W)$, which is a diagonal matrix, with its diagonal
 elements corresponding to those of $W,$  and $W_2=W-W_1.$
%Denote $A_{ij}^+=E_{ij}^+$~ if~ $E_{ij}^+>0;$ otherwise 0.     $A_{ij}^-=E_{ij}^-$~ if~ $E_{ij}^->0;$ otherwise 0.

  According to Chen ${\it et ~ al.}$ (\cite{chen2014overlapping}),
  the goal is to maximize the logarithm of the likelihood of the signed network below with respect to the unknown  parameters $W$ and the community membership matrix $H$.
\begin{eqnarray*}%\label{signed_1}
  &&\max_{H,\, W}\log P(\mathscr{G}|{H}, W)\\ &&=  \max_{{H},\, W}\sum_{{i,\,j=1}}^n \left[{A^+_{ij}}\log\left(\sum_{r,\,r}w_{rr}H_{ir}H_{jr}\right)  +{A^-_{ij}}\log\left(\sum_{r,\,s\,(r\neq s)}w_{rs}H_{ir}H_{js} \right)\right],
\end{eqnarray*}
which is equivalent to
\begin{eqnarray}\label{signed_2}
 \min_{H,\, W}\sum_{{i,\,j=1}}^n \left({A^+_{ij}}\log\frac{1}{\displaystyle\sum_{r,\,r}w_{rr}H_{ir}H_{js}}  +{A^-_{ij}}\log\frac{1}{\displaystyle\sum_{r,\,s\,(r\neq s)}w_{rs}H_{ir}H_{js} }\right).
\end{eqnarray}
Firstly,
\begin{eqnarray}\label{signed_3}
&&\sum_{{i,\,j=1}}^n \left({A^+_{ij}}\log\frac{1}{\displaystyle\sum_{r,\,r}w_{rr}H_{ir}H_{js}}  +{A^-_{ij}}\log\frac{1}{\displaystyle\sum_{r,\,s\,(r\neq s)}w_{rs}H_{ir}H_{js} }\right)\nonumber \\
% &&=\sum_{{i,j=1}}^n \left\{A_{ij}^+\log\frac{1}{\sum_{rr}w_{rr}H_{ir}H_{js}}+\sum_{i,j}A_{ij}^-\log\frac{1}{\sum_{rs(r\neq s)}w_{rs}H_{ir}H_{js}}\right\}\nonumber\\
&&=\sum_{{i,\,j=1}}^n\left[   A_{ij}^+\log\frac{1}{(H W_1 H^T)_{ij}}+ A_{ij}^- \log\frac{1}{ (H W_2 H^T)_{ij} } \right].
\end{eqnarray}

Also, note that
\begin{eqnarray}\label{signed_4}
\sum_{i,\,j}(H W H^T)_{ij}&=&\sum_{i,\,j}\sum_{r,\,s} w_{rs}H_{ir}H_{js}\nonumber\\
&=&\sum_{r,\,s}w_{rs}\sum_iH_{ir}\sum_jH_{js}=\sum_{r,\,s}w_{rs}=1,
\end{eqnarray}
and on the other side,
\begin{eqnarray}\label{signed_5}
\sum_{i,\,j}(H W H^T)_{ij}&=&\sum_{i,\,j}\left(H(W_1+W_2) H^T\right)_{ij}\nonumber\\
&=&\sum_{i,\,j}(H W_1 H^T)_{ij}+\sum_{i,\,j}(H W_2 H^T)_{ij}.
\end{eqnarray}

By combining  (\ref{signed_3})  (\ref{signed_4}) and (\ref{signed_5}), and adding the constraints on $H$ and $W$, we have that the optimal problem (\ref{signed_2}) is equivalent to the following joint non-negative matrix factorization model:
%\begin{eqnarray}
%  \min_{H, W}&& \sum_{i,j} A_{ij}^+\log\frac{1}{(H W_1 H^T)_{ij}}
%+\sum_{i,j}A_{ij}^-\log\frac{1}{{(H W_2 H^T})_{ij}}\nonumber \\
% &&+\sum_{i,j}(H W_1 H^T)_{ij}+\sum_{i,j}(H W_2 H^T)_{ij}-\sum_{i,j}(A^+_{ij}+A^-_{ij}).
%\end{eqnarray}

%Remembering the constraint conditions, thus optimal problem (\ref{signed_2}) has an  equivalent joint non-negative matrix factorization form below

$$\left\{
\begin{array}{rl}\vspace{5mm}
 \displaystyle\min_{H,\, W} & \left\{\displaystyle\sum_{i,\,j=1}^n\left[ A_{ij}^+\log\frac{1}{{(H W_1 H^T})_{ij}}+(H W_1 H^T)_{ij}-A^+_{ij}\right]\right.\\\vspace{5mm}
 & \left.+\displaystyle\sum_{i,\,j=1}^n\left[A_{ij}^-\log\frac{1}{{(H W_2 H^T})_{ij}}+(H W_2 H^T)_{ij}-A^-_{ij}\right]\right\}  \\
 s.t. &\displaystyle\sum_{i=1}^n H_{ir}=1,~ \sum_{r,\,s} (W_1+W_2)_{rs}=1.\\\\
  & W_1 \textrm{ is nonnegative and diagonal, }\\\\
  & W_2\textrm{ is nonnegative and its diagonal elements are zeros.}\\\\
  & H_{ir}\geq0, i=1,2,\cdots,n; r=1,2,\cdots,c.
\end{array}
\right.$$

%From the four cases of blockmodels, we obtain the equivalence of blockmodels and non-negative matrix factorization.

\section{Experimental Results}\label{experi}
Although the likelihood function of SBM can be reformulated as the objective function of NMF, their algorithms are different. In this section, we use synthetic networks to compare the effectiveness of the algorithms employed by SBM and NMF, respectively.
\subsection{Algorithms for SBM and NMF}
There is a package ``blockmodels'' (\cite{leger2016R}) in \textbf{\textsf{R}}, which uses variational EM algorithm to estimate the parameters in SBM with some common probability distribution functions including Bernoulli distribution, Poisson distribution and Gaussian distribution (\cite{mariadassou2010uncovering}), and explore the community number by the ICL criterion (\cite{biernacki2000assessing}). We use the command \texttt{BM\_poisson} in the package, and fix the community number by setting both the parameters \texttt{explore\_min} and \texttt{explore\_max} to be the true community number.

We designed the multiplicative update rules for the nonnegative matrix factorization model (\ref{2eq2}) and (\ref{normal1}) (\cite{lee2001algorithms, ding2005equivalence}), which are summarized in Algorithm \ref{Al:01} and Algorithm \ref{Al:02}, respectively.
 We set the iteration number \texttt{iter} to 500 for each of the algorithms.
\renewcommand{\algorithmicrequire}{\textbf{Input:}}
\renewcommand{\algorithmicensure}{\textbf{Output:}}
\begin{algorithm}[H]
\caption{NMF with Kullback-Leibler Divergence, model (\ref{2eq2})}
\label{Al:01}
\begin{algorithmic}[1]
\REQUIRE $A,$ iter
\ENSURE $G,W$
\FOR{$t=1:\mbox{iter}$}\vspace{2mm}
\STATE
$
\displaystyle G_{ij}:=G_{ij}\frac{\Big(\displaystyle\frac{A}{GWG^T}GW\Big)_{ij}}{\sum\limits_l(WG^T)_{jl}}$\vspace{2mm}
\STATE $\displaystyle \underline{W_{ij}:=W_{ij}\frac{\Big(G^T\displaystyle\frac{A}{(GWG^T)}G\Big)_{ij}}
           {\sum\limits_{k,\,l}G_{ki}G_{lj}}}
$\vspace{2mm}
\STATE
$
\displaystyle{G_{ij}:=\frac{G_{ij}}{\sum\limits_jG_{ij}}}\vspace{2mm}
$
% \STATE
%$
%\displaystyle G_{ik}:=G_{ik}\frac{(X^TF)_{ik}}{(GF^{T}F)_{ik}}
%$
\ENDFOR
\end{algorithmic}
\end{algorithm}

\renewcommand{\algorithmicrequire}{\textbf{Input:}}
\renewcommand{\algorithmicensure}{\textbf{Output:}}
\begin{algorithm}[H]
\caption{NMF with Least Squares Error, model (\ref{normal1})}
\label{Al:02}
\begin{algorithmic}[1]
\REQUIRE $A,$ iter
\ENSURE $U$
\FOR{$t=1:\mbox{iter}$}\vspace{2mm}
\STATE
$\vspace{2mm}
\displaystyle G_{ij}:=G_{ij}\frac{(AGW)_{ij}}{(GWG^TGW)_{ij}}$
\STATE $\displaystyle \underline{W_{ij}:=W_{ij}\frac{(G^TAG)_{ij}}{(G^TGWG^TG)_{ij}}}
$\vspace{2mm}
\STATE
$
\displaystyle{G_{ij}:=\frac{G_{ij}}{\sum\limits_jG_{ij}}}\vspace{2mm}
$
% \STATE
%$
%\displaystyle G_{ik}:=G_{ik}\frac{(X^TF)_{ik}}{(GF^{T}F)_{ik}}
%$
\ENDFOR
\end{algorithmic}
\end{algorithm}

\subsection {Datasets Description}
In this paper we use the computer-generated networks for comparison.
\begin{enumerate}
\item The Girvan-Newman benchmark network (GN, \cite{girvan2002community}). The GN network contains four communities with 32 vertices each.
On average, the number of edges between two vertices from the same community is $Z_{in}$ , and that from different communities is $Z_{out}$.
 As expected, the communities become less  clear
as $Z_{out}$  increases. Here $Z_{in} + Z_{out}$ is set to be 16.

\item The Lancichinetti-Fortunato-Radicchi benchmark network (LFR, \cite{lancichinetti2009benchmarks}).
The LFR network was proposed to cover most
characteristics of real networks, such as size of the network and
heterogeneous degree distribution, which the GN networks did
not capture. In LFR benchmarks, distributions of  both the degree and the
community size
 obey power laws with exponents
$\alpha$ and $\beta$ respectively.  With probability $\mu$, a vertex connects to another vertex from different communities, and
in its own community with probability $1-\mu$.

In this paper,  the parameters of the LFR benchmark are set  as
follows: The number of vertices is 1000, the maximum of degree is
50, the exponents are $\alpha=$2, $\beta=1$, the average
degree of the nodes is 20 and the range of the mixing parameter
$p$ is from 0.1 to 0.9.
\end{enumerate}

\subsection{Simulation Results}
In this subsection, we compare the numerical results of the algorithms employed by SBM and NMF on GN and LFR networks. We use the normalized mutual information (NMI, \cite{strehl2003cluster}) to evaluate the quality of the results, which can be formulated as follows:
$$\displaystyle
  I(M_1, M_2) = \frac{\sum\limits_{i=1}^{k}
     \sum\limits_{j=1}^{k}n_{ij}\log\displaystyle\frac{n_{ij}n}{n_i^{(1)}n_j^{(2)}}}
  {\sqrt{\left(\sum\limits_{i=1}^{k}n_i^{(1)}\log\displaystyle\frac{n_i^{(1)}}{n}\right)
  \left(\sum\limits_{j=1}^{k}n_j^{(2)}\log\displaystyle\frac{n_j^{(2)}}{n}\right)}},
$$
where $M_1$ and $M_2$ are the implanted community label and the computed community label, respectively; $k$ is the true community number; $n$ is the number of nodes; $n_{ij}$ is the number of nodes in the implanted community $i$ that are assigned to the computed community $j$; $n_i^{(1)}$ is the number of nodes in the implanted community $i$;  $n_j^{(2)}$ is the number of nodes in the computed community $j$; and $\log$ here is the natural logarithm. The larger the NMI value, the better the community partition.

\begin{figure}
\hspace{-7mm}
\includegraphics[height=50mm,width=160mm]{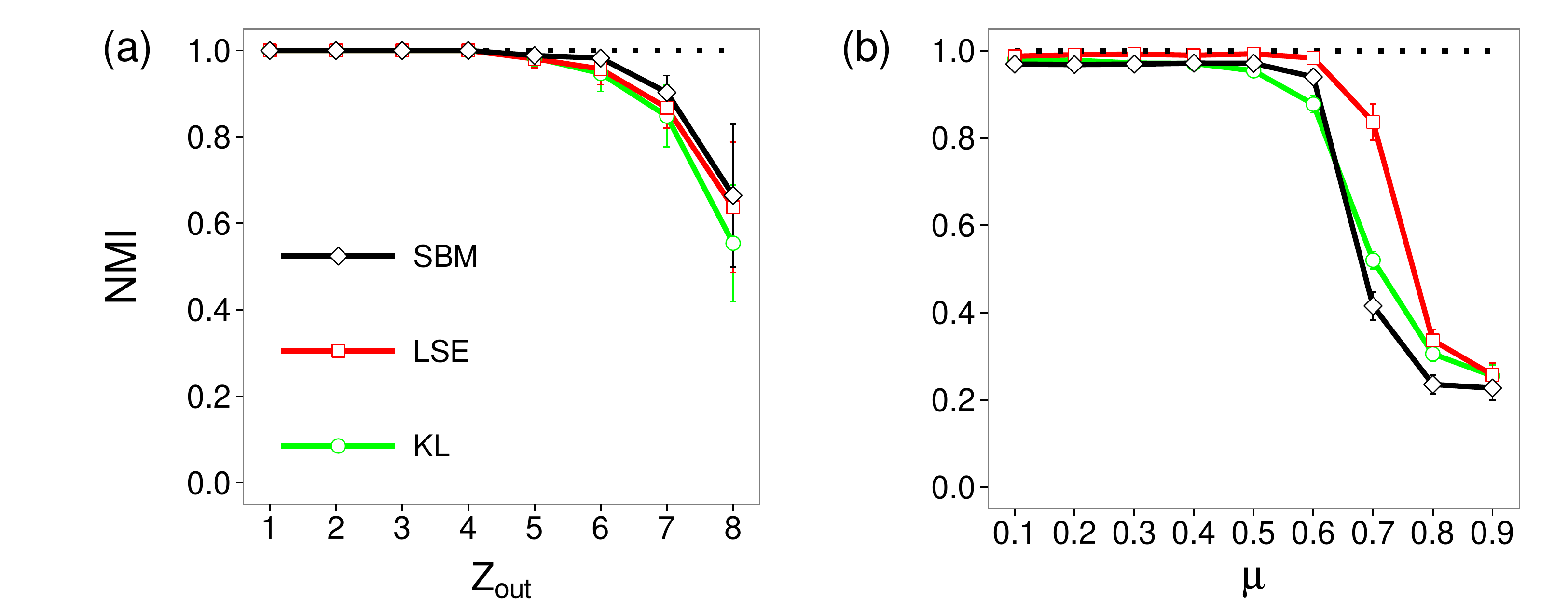}
\caption{Averaged NMI with the standard deviation of SBM and NMF on (a) GN networks and (b) LFR networks. SBM stands for stochastic block model, LSE stands for nonnegative matrix factorization with Least Square Error, and KL stands for nonnegative matrix factorization with KL divergence.}\label{Fig:01}
\end{figure}

The results are averaged over ten trials and are shown in Fig. \ref{Fig:01}.
From the figure,  one can conclude the following:
(1) The algorithms employed by SBM and NMF are different, although their models to be optimized are equivalent.
(2) There is no single winner. SBM works slightly better when the community structures are clear, and NMF with Least Square Error performs better when the degree heterogeneity is included, especially when the community structures are fuzzy.
\section{Conclusion}\label{conclu}
In this paper, we give the detailed analysis on the connections between likelihood maximization of different stochastic block models and nonnegative matrix factorization. The studied stochastic block models include the standard SBM, the degree-corrected SBM, the bipartite SBM, the normal distributed SBM, the directed SBM, and the signed SBM. Preliminary  numerical experiments are also performed on synthetic networks to compare the difference between the algorithms for SBM and NMF.

The connections have several immediate implications: Firstly, it provides a rigorous statistical interpretation of NMF. NMF has been successfully applied to community detection, but it is still not very clear that why the model works. The equivalence results give us a preliminary interpretation. Secondly, an interesting point is that the principled statistical models can be reduced to optimization problems, making it possible to design algorithms from different perspectives. Finally, the relations among SBM, NMF and modularity maximization is an interesting problem for future work, since it is proved that the degree corrected SBM model is equivalent with modularity maximization recently \cite{newman2016community}.
Other interesting problems for future work include the general relations between the generative models and NMF, a systematically comparison among the algorithms employed by SBM and NMF, how to combine the algorithms to make them profit from each other and make up for each other's deficiencies.

\section*{References}
\bibliographystyle{elsarticle-num}
\bibliography{sbm}

\begin{thebibliography}{10}
\expandafter\ifx\csname url\endcsname\relax
  \def\url#1{\texttt{#1}}\fi
\expandafter\ifx\csname urlprefix\endcsname\relax\def\urlprefix{URL }\fi
\expandafter\ifx\csname href\endcsname\relax
  \def\href#1#2{#2} \def\path#1{#1}\fi

\bibitem{girvan2002community}
M.~Girvan, M.~E. Newman, Community structure in social and biological networks,
  Proceedings of the national academy of sciences 99~(12) (2002) 7821--7826.

\bibitem{newman2004finding}
M.~E. Newman, M.~Girvan, Finding and evaluating community structure in
  networks, Physical Review E 69~(2) (2004) 026113.

\bibitem{nematzadeh2014optimal}
A.~Nematzadeh, E.~Ferrara, A.~Flammini, Y.-Y. Ahn, Optimal network modularity
  for information diffusion, Physical review letters 113~(8) (2014) 088701.

\bibitem{holland1983stochastic}
P.~W. Holland, K.~B. Laskey, S.~Leinhardt, Stochastic blockmodels: First steps,
  Social networks 5~(2) (1983) 109--137.

\bibitem{snijders1997estimation}
T.~A. Snijders, K.~Nowicki, Estimation and prediction for stochastic
  blockmodels for graphs with latent block structure, Journal of classification
  14~(1) (1997) 75--100.

\bibitem{nowicki2001estimation}
K.~Nowicki, T.~A.~B. Snijders, Estimation and prediction for stochastic
  blockstructures, Journal of the American Statistical Association 96~(455)
  (2001) 1077--1087.

\bibitem{newman2007mixture}
M.~E. Newman, E.~A. Leicht, Mixture models and exploratory analysis in
  networks, Proceedings of the National Academy of Sciences 104~(23) (2007)
  9564--9569.

\bibitem{goldenberg2010survey}
A.~Goldenberg, A.~X. Zheng, S.~E. Fienberg, E.~M. Airoldi, A survey of
  statistical network models, Foundations and Trends{\textregistered} in
  Machine Learning 2~(2) (2010) 129--233.

\bibitem{airoldi2009mixed}
E.~M. Airoldi, D.~M. Blei, S.~E. Fienberg, E.~P. Xing, Mixed membership
  stochastic blockmodels, in: Advances in Neural Information Processing
  Systems, 2009, pp. 33--40.

\bibitem{bickel2009nonparametric}
P.~J. Bickel, A.~Chen, A nonparametric view of network models and
  newman--girvan and other modularities, Proceedings of the National Academy of
  Sciences 106~(50) (2009) 21068--21073.

\bibitem{newman2012communities}
M.~E. Newman, Communities, modules and large-scale structure in networks,
  Nature Physics 8~(1) (2012) 25--31.

\bibitem{karrer2011stochastic}
B.~Karrer, M.~E. Newman, Stochastic blockmodels and community structure in
  networks, Physical Review E 83~(1) (2011) 016107.

\bibitem{decelle2011asymptotic}
A.~Decelle, F.~Krzakala, C.~Moore, L.~Zdeborov{\'a}, Asymptotic analysis of the
  stochastic block model for modular networks and its algorithmic applications,
  Physical Review E 84~(6) (2011) 066106.

\bibitem{aicher2014learning}
C.~Aicher, A.~Z. Jacobs, A.~Clauset, Learning latent block structure in
  weighted networks, Journal of Complex Networks (2014) cnu026.

\bibitem{shen2011exploring}
H.-W. Shen, X.-Q. Cheng, J.-F. Guo, Exploring the structural regularities in
  networks, Physical Review E 84~(5) (2011) 056111.

\bibitem{chen2014overlapping}
Y.~Chen, X.~Wang, B.~Yuan, B.~Tang, Overlapping community detection in networks
  with positive and negative links, Journal of Statistical Mechanics: Theory
  and Experiment 2014~(3) (2014) P03021.

\bibitem{larremore2014efficiently}
D.~B. Larremore, A.~Clauset, A.~Z. Jacobs, Efficiently inferring community
  structure in bipartite networks, Physical Review E 90~(1) (2014) 012805.

\bibitem{schaeffer2007graph}
S.~E. Schaeffer, Graph clustering, Computer Science Review 1~(1) (2007) 27--64.

\bibitem{krzakala2013spectral}
F.~Krzakala, C.~Moore, E.~Mossel, J.~Neeman, A.~Sly, L.~Zdeborov{\'a},
  P.~Zhang, Spectral redemption in clustering sparse networks, Proceedings of
  the National Academy of Sciences 110~(52) (2013) 20935--20940.

\bibitem{newman2006modularity}
M.~E. Newman, Modularity and community structure in networks, Proceedings of
  the national academy of sciences 103~(23) (2006) 8577--8582.

\bibitem{lee1999learning}
D.~D. Lee, H.~S. Seung, Learning the parts of objects by non-negative matrix
  factorization, Nature 401~(6755) (1999) 788--791.

\bibitem{lee2001algorithms}
D.~D. Lee, H.~S. Seung, Algorithms for non-negative matrix factorization, in:
  Advances in neural information processing systems, 2001, pp. 556--562.

\bibitem{li2001learning}
S.~Z. Li, X.~W. Hou, H.~Zhang, Q.~Cheng, Learning spatially localized,
  parts-based representation, in: Computer Vision and Pattern Recognition,
  2001. CVPR 2001. Proceedings of the 2001 IEEE Computer Society Conference on,
  Vol.~1, IEEE, 2001, pp. I--207.

\bibitem{cooper2002summarizing}
M.~Cooper, J.~Foote, Summarizing video using non-negative similarity matrix
  factorization, in: Multimedia Signal Processing, 2002 IEEE Workshop on, IEEE,
  2002, pp. 25--28.

\bibitem{brunet2004metagenes}
J.-P. Brunet, P.~Tamayo, T.~R. Golub, J.~P. Mesirov, Metagenes and molecular
  pattern discovery using matrix factorization, Proceedings of the national
  academy of sciences 101~(12) (2004) 4164--4169.

\bibitem{pauca2004text}
V.~P. Pauca, F.~Shahnaz, M.~W. Berry, R.~J. Plemmons, Text mining using
  non-negative matrix factorizations., in: SDM, Vol.~4, SIAM, 2004, pp.
  452--456.

\bibitem{sra2005generalized}
S.~Sra, I.~S. Dhillon, Generalized nonnegative matrix approximations with
  bregman divergences, in: Advances in neural information processing systems,
  2005, pp. 283--290.

\bibitem{shashua2005non}
A.~Shashua, T.~Hazan, Non-negative tensor factorization with applications to
  statistics and computer vision, in: Proceedings of the 22nd international
  conference on Machine learning, ACM, 2005, pp. 792--799.

\bibitem{zarei2009detecting}
M.~Zarei, D.~Izadi, K.~A. Samani, Detecting overlapping community structure of
  networks based on vertex--vertex correlations, Journal of Statistical
  Mechanics: Theory and Experiment 2009~(11) (2009) P11013.

\bibitem{psorakis2011overlapping}
I.~Psorakis, S.~Roberts, M.~Ebden, B.~Sheldon, Overlapping community detection
  using bayesian non-negative matrix factorization, Physical Review E 83~(6)
  (2011) 066114.

\bibitem{zhang2013overlapping}
Z.-Y. Zhang, Y.~Wang, Y.-Y. Ahn, Overlapping community detection in complex
  networks using symmetric binary matrix factorization, Physical Review E
  87~(6) (2013) 062803.

\bibitem{cao2013identifying}
X.~Cao, X.~Wang, D.~Jin, Y.~Cao, D.~He, Identifying overlapping communities as
  well as hubs and outliers via nonnegative matrix factorization, Scientific
  reports 3.

\bibitem{zhang2015community}
Z.-Y. Zhang, Y.-Y. Ahn, Community detection in bipartite networks using
  weighted symmetric binary matrix factorization, International Journal of
  Modern Physics C 26~(09) (2015) 1550096.

\bibitem{ding2005equivalence}
C.~H. Ding, X.~He, H.~D. Simon, On the equivalence of nonnegative matrix
  factorization and spectral clustering, in: SDM, Vol.~5, SIAM, 2005, pp.
  606--610.

\bibitem{ding2008equivalence}
C.~Ding, T.~Li, W.~Peng, On the equivalence between non-negative matrix
  factorization and probabilistic latent semantic indexing, Computational
  Statistics \& Data Analysis 52~(8) (2008) 3913--3927.

\bibitem{leger2016R}
J.-B. Leger, Blockmodels: A {R}-package for estimating in latent block model
  and stochastic block model, with various probability functions, with or
  without covariates (2016).
\newblock \href {http://arxiv.org/abs/1602.07587} {\path{arXiv:1602.07587}}.

\bibitem{mariadassou2010uncovering}
M.~Mariadassou, S.~Robin, C.~Vacher, Uncovering latent structure in valued
  graphs: a variational approach, The Annals of Applied Statistics (2010)
  715--742.

\bibitem{biernacki2000assessing}
C.~Biernacki, G.~Celeux, G.~Govaert, Assessing a mixture model for clustering
  with the integrated completed likelihood, Pattern Analysis and Machine
  Intelligence, IEEE Transactions on 22~(7) (2000) 719--725.

\bibitem{lancichinetti2009benchmarks}
A.~Lancichinetti, S.~Fortunato, Benchmarks for testing community detection
  algorithms on directed and weighted graphs with overlapping communities,
  Physical Review E 80~(1) (2009) 016118.

\bibitem{strehl2003cluster}
A.~Strehl, J.~Ghosh, Cluster ensembles---a knowledge reuse framework for
  combining multiple partitions, The Journal of Machine Learning Research 3
  (2003) 583--617.

\bibitem{newman2016community}
M.~Newman, Community detection in networks: Modularity optimization and maximum
  likelihood are equivalent, arXiv preprint arXiv:1606.02319.

\end{thebibliography}

\end{document}